\documentclass[aps, prx, 10pt, twocolumn,floatfix, nofootinbib]{revtex4-2}
\usepackage{amsmath,amssymb,amsfonts,bm}
\usepackage[breaklinks=true,colorlinks,allcolors=blue]{hyperref}
\usepackage{graphicx}
\usepackage{xcolor}
\usepackage{txfonts}
\usepackage{braket}
\usepackage{comment}
\usepackage{siunitx}
\usepackage{enumitem}
\usepackage{mathtools}
\usepackage[normalem]{ulem}
\usepackage[dvipsnames]{xcolor}
\usepackage{tabularx, booktabs, multirow}
\usepackage[table]{xcolor}
\usepackage{orcidlink}
\usepackage{physics}
\usepackage{lipsum}
\usepackage{booktabs}
\usepackage[ruled,vlined,linesnumbered]{algorithm2e}
\usepackage{mathrsfs}

\newcommand{\alglabel}[1]{Alg.~\ref{#1}}

\newcommand{\seclabel}[1]{Sec.~\ref{#1}}
\newcommand{\tablabel}[1]{Table~\ref{#1}}

\begin{document}
\title{Large-scale portfolio optimization on a trapped-ion quantum computer} 

\author{Alejandro Gomez Cadavid$^{\orcidlink{0000-0003-3271-4684}1,2}$}

\author{Ananth Kaushik$^{\orcidlink{0009-0009-2799-1194} 7}$}

\author{Pranav Chandarana$^{\orcidlink{}1,2}$}

\author{Miguel Angel Lopez-Ruiz$^{\orcidlink{0000-0002-8152-5655} 7}$}

\author{Gaurav Dev$^{\orcidlink{}1,3}$}

\author{Willie Aboumrad$^{\orcidlink{} 7}$}

\author{Qi Zhang$^{\orcidlink{0000-0001-6223-5516}1}$}

\author{Claudio Girotto$^{\orcidlink{0000-0001-8739-5866} 7}$}

\author{Sebastián V. Romero$^{\orcidlink{0000-0002-4675-4452}1,4,5}$}

\author{Martin Roetteler$^{\orcidlink{0000-0003-0234-2496} 7}$}

\author{Enrique Solano$^{\orcidlink{0000-0002-8602-1181}1}$}
\email{enr.solano@gmail.com}

\author{Marco Pistoia$^{\orcidlink{0000-0001-9002-1128} 7, 8}$}

\author{Narendra N. Hegade$^{\orcidlink{0000-0002-9673-2833}1,6}$}
\email{narendrahegade5@gmail.com}

\affiliation{\medskip $^{1}$Kipu Quantum GmbH, Greifswalderstrasse 212, 10405 Berlin, Germany\\$^{2}$\mbox{Department of Physical Chemistry, University of the Basque Country EHU, Apartado 644, 48080 Bilbao, Spain}\\$^{3}$\mbox{TUM School of Management, Technische Universität München, Bildungscampus 9, 74076 Heilbronn, Germany}\\$^{4}$\mbox{Instituto de Ciencia de Materiales de Madrid (CSIC), Cantoblanco, E-28049 Madrid, Spain}\\$^{5}$\mbox{Departamento de Física Teórica de la Materia Condensada, Universidad Autónoma de Madrid, E-28049 Madrid, Spain}\\$^{6}$\mbox{IDAL, Electronic Engineering Department, ETSE-UV, University of Valencia, Avgda. Universitat s/n, 46100 Burjassot, Valencia, Spain}\\$^{7}$IonQ Inc., 4505 Campus Dr, College Park, MD 20740, USA\\$^{8}$IonQ Italia, Rome, Italy}

\date{\today}
\begin{abstract}
We present an end-to-end pipeline for large-scale portfolio selection with cardinality constraints and experimentally demonstrate it on trapped-ion quantum processors using hardware-aware decomposition. Building on RMT-based correlation-matrix denoising and community detection, we identify correlated asset groups and introduce a correlation-guided greedy splitting scheme that caps each cluster by the executable qubit budget. Each cluster defines a hardware-embeddable QUBO subproblem that we solve using bias-field digitized counterdiabatic quantum optimization (BF-DCQO), a non-variational method that avoids classical parameter-training loops. We recombine low-energy candidates into global portfolios and enforce feasibility with a two-stage post-processing routine: fast repair followed by a cardinality-preserving swap local search. We benchmark the workflow on a 250-asset universe taken from the S\&P 500 and execute subproblems on a 64-qubit Barium development system similar to the forthcoming IonQ Tempo line.
We observe that larger executable subproblem sizes reduce decomposition error and systematically improve final objective values and risk–return trade-offs relative to randomized baselines under identical post-processing. Overall, the results establish a hardware-tested route for scaling financial optimization problems, defined by a trade space in which executable problem size and circuit cost are balanced against the resulting solution quality.
\end{abstract}

\maketitle

\section{Introduction}

Optimization problems arising in financial decision-making often involve discrete choices, combinatorial constraints, and noisy input data~\cite{mansini2014twenty, ceria2006incorporating}. Among them, portfolio selection with cardinality constraints is a prototypical challenging problem~\cite{kellerer2000, CHANG20001271,jobst2001computational, bertsimas2009algorithm}, in which an investor must choose a fixed number of assets while balancing expected return against risk, building upon the classical mean-variance framework of Markowitz~\cite{markowitz1952}. As problem sizes increase, classical methods quickly become computationally impractical, and heuristic approaches tend to dominate in realistic settings.

Quantum computing has emerged as a promising complementary technology for combinatorial optimization. Many financial formulations can be expressed as quadratic unconstrained binary optimization (QUBO) problems and mapped to Ising models~\cite{lucas2014ising, bertsimas2009algorithm}, enabling the use of quantum algorithms such as quantum annealing \cite{rosenberg2015solving, albash2018}, the quantum approximate optimization algorithm (QAOA)~\cite{farhi2014}, and digitized counterdiabatic approaches such as digitized counterdiabatic quantum optimization (DCQO)~\cite{Hegade_2021, PhysRevResearch.4.043204} and bias-field DCQO (BF-DCQO)~\cite{cadavid2025, Romero2025}. 
However, current quantum processors remain constrained by limited qubit counts, restricted connectivity, and finite coherence times, which collectively bound the size and complexity of executable problem instances. These practical limitations motivate the development of frameworks that explicitly incorporate hardware constraints into the design of quantum optimization algorithms, enabling more scalable and reliable implementations on contemporary quantum platforms ~\cite{egger2021,harrigan2021,pelofske2024}. In this regard, several efforts have sought to tackle large-scale industrial problems on current quantum computers, either by decomposing problems into subproblems or by encoding them in a qubit-efficient manner~\cite{zhao2022hybrid, PhysRevResearch.7.023142, soloviev2025large}.

Existing demonstrations of quantum optimization for financial applications are constrained by device size and by the practical cost of executing strongly coupled objective functions under near-term noise and runtime limitations. Cardinality-constrained portfolio selection is a representative example: it is NP-hard, induces many pairwise couplings, and requires enforcing an exact $K$-asset constraint, all of which complicate direct hardware execution at realistic scales. Recent work has shown that correlation-structure-aware decomposition can mitigate these challenges by denoising empirical correlation matrices and partitioning the asset universe into communities that can be solved separately~\cite{PhysRevResearch.7.023142}.

We build on this direction with an end-to-end workflow that explicitly ties the decomposition to the executable qubit budget and instantiates the solve stage with a non-variational quantum optimizer. Starting from a QUBO/Ising formulation of fixed-cardinality portfolio selection, we enforce a strict hardware size cap via a correlation-guided splitting rule that converts oversized communities into clusters that fit a target qubit budget. We solve the resulting Ising subproblems on trapped-ion quantum hardware using bias-field digitized counterdiabatic quantum optimization (BF-DCQO)~\cite{Hegade_2021,cadavid2025,Romero2025}, and construct global portfolios by recombining low-energy subproblem candidates followed by a cardinality-enforcing refinement step.

We benchmark the approach on a 250-asset universe and experimentally execute subproblems with 36 qubits and up to 60 qubits on a 64-qubit trapped-ion system. We observe that larger executable subproblem sizes reduce decomposition error and improve final objective values and risk-return trade-offs relative to classical cluster-based and randomized baselines under identical post-processing. This establishes a direct link between available qubit budget, decomposition granularity, and end-to-end solution quality, supporting hardware-aware quantum optimization workflows for applied financial decision-making on near-term devices.

The remainder of this paper is organized as follows. In Sec.~\ref{sec:portfolio_optimization} we present the fixed-cardinality portfolio selection problem and the QUBO/Ising mapping used throughout. In Sec.~\ref{sec:methods} we detail the hardware-aware pipeline, including RMT-based denoising of correlations, community detection with an explicit qubit-budget cap, BF-DCQO execution on the resulting subproblems, and a cardinality-preserving local-search refinement used at both the cluster and global levels. In Sec.~\ref{sec:experimental_results} we evaluate the pipeline on a 250-asset universe and report trapped-ion hardware executions on IonQ systems with 36-qubit and 64-qubit configurations (executing subproblems up to 60 qubits), highlighting how decomposition granularity and reconstructed portfolio quality change with available qubits. We conclude with implications for near-term hardware execution of dense Ising/QUBO objectives and extensions enabled by larger qubit counts and improved fidelities.

\section{Portfolio Optimization Problem}
\label{sec:portfolio_optimization}

Portfolio selection aims to identify an optimal subset of assets that balances expected return and risk, subject to practical constraints such as cardinality, diversification, and regulatory requirements. In this work, we consider a combinatorial portfolio optimization problem in which decisions are binary: each asset \(i\) is either included (\(x_i = 1\)) or excluded (\(x_i = 0\)) from the portfolio. This discrete structure allows a direct mapping to quadratic unconstrained binary optimization (QUBO) objectives and Ising Hamiltonians compatible with quantum optimization algorithms such as BF-DCQO.

\textit{Problem definition:} Let \(\mu_i\) denote the expected return of asset \(i\), and let \(C \in \mathbb{R}^{n \times n}\) be the covariance matrix. The goal is to select a subset of exactly \(K\) assets (a fixed-cardinality constraint) that optimizes a risk--return trade-off. The binary vector \(x = (x_1,\dots,x_n)\) must satisfy \(\sum_{i=1}^{n} x_i = K\). The classical objective
\begin{equation}
\label{eq:min-problem}
\begin{aligned}
\min_{x\in\{0,1\}^{n}} \;& \frac{\gamma-1}{2}\mu^T x + \frac{\gamma+1}{2} x^T Cx \\
\text{s.t.} \;& \sum_{i=1}^{n} x_i = K,
\end{aligned}
\end{equation}
combines a risk term \(\sum_{i,j} C_{ij} x_i x_j\) with a linear return term \(\sum_i \mu_i x_i\), using a weighting parameter \(\gamma \in [-1,1]\) to interpolate between pure risk-minimization (\(\gamma=1\)) and pure return-maximization (\(\gamma=-1\)). In this paper, we set \(\gamma=0\) to balance the risk and return terms in Eq.~\eqref{eq:min-problem}.

\textit{QUBO formulation:} To encode the cardinality constraint into a QUBO, we augment the objective with a quadratic penalty of the form \((\sum_{i=1}^{n} x_i - K)^2\), multiplied by a sufficiently large coefficient to enforce feasibility. After expanding and collecting terms, the full cost function can be written in the standard QUBO form
\begin{equation}
\label{eq:cost_fn}
f(x) = x^\top Q x + q^\top x + \text{const},
\end{equation}
where \(Q\) contains contributions from the covariance term and the quadratic part of the penalty, and \(q\) collects the linear return terms and the linear part of the penalty. This representation is directly compatible with classical as well as with quantum heuristics. See Appendix~\ref{appendix:ising-qubo-formulation} for the explicit construction of \(Q\) and \(q\) from Eq.~\eqref{eq:min-problem}.

\textit{Ising representation:} For quantum hardware that natively implements Pauli interactions, it is convenient to convert binary variables to Ising spins via \(x_i = (1 - z_i)/2\), where \(z_i \in \{-1, +1\}\). Substitution yields an Ising Hamiltonian of the generic form
\begin{equation}
\label{eq:cost_ising}
H_{f} = \sum_i h_i z_i + \sum_{i<j} J_{ij} z_i z_j + \mathrm{const.},
\end{equation}
with fields \(h_i\) and couplings \(J_{ij}\) obtained through a linear transformation of \(Q\) and \(q\), see Appendix~\ref{appendix:ising-qubo-formulation}. This structure aligns directly with the BF-DCQO optimizer described in Sec.~\ref{sec:quantum-optimizer}, where \(H_f\) is the target Hamiltonian.

\begin{figure*}
    \centering
    \includegraphics[width=1\linewidth]{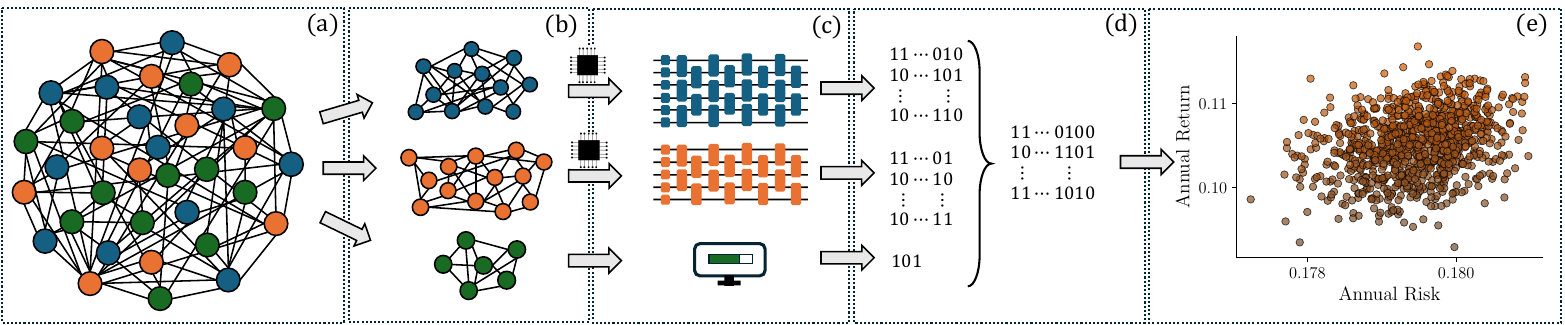}
    \caption{Schematic of the end-to-end pipeline: (a) original problem; (b) community detection and size-bounded splitting produce clusters with \(|\mathcal{C}_m|\le Q_{\max}\); (c) each cluster is mapped to an Ising instance and solved via BF-DCQO (or classically for small clusters); (d--e) low-energy cluster candidates are recombined into global portfolios and refined using a cardinality-preserving local search.}
    \label{fig:schematics}
\end{figure*}

\section{Methods}
\label{sec:methods}

\subsection{Hardware-aware decomposition pipeline}
\label{sec:methods-Hardware-aware-decomposition-pipeline}

To ensure that each QUBO subproblem fits on a target quantum processor with a maximum qubit capacity $Q_{\max}$, we partition the asset universe into clusters whose sizes do not exceed this limit. This is achieved through \alglabel{alg:get_clusters}. The method proceeds in two stages. First, we apply a denoising decomposition of the correlation matrix $\mathscr{C}$ based on Refs.~\cite{PhysRevResearch.7.023142, dcmppln2024}. where the correlation matrix is partitioned into three spectral components:
\begin{equation}
    \mathscr{C} = \mathscr{C}_{\text{noise}} + \mathscr{C}_{\star} + \mathscr{C}_{\text{global}},
\end{equation}
where $\mathscr{C}_{\text{noise}}$ corresponds to eigenmodes consistent with noise, $\mathscr{C}_{\text{global}}$ is the dominant market mode, and $\mathscr{C}_{\star}$ contains the structured, informative eigenmodes. Clustering is performed only on $\mathscr{C}_{\star}$, which improves robustness by suppressing noise-driven relationships. 

A community detection method is applied to $\mathscr{C}_{\star}$, producing an initial hard partition of assets. These communities reflect meaningful correlation structure, but their sizes may exceed the hardware-bound $Q_{\max}$. Therefore, a second stage enforces the maximum cluster size without discarding any assets. Oversized communities are recursively split using a correlation-aware greedy procedure: within each community, we select a high-degree node (in the absolute-correlation graph) and group it with its most strongly correlated neighbours up to size $Q_{\max}$. The selected nodes form a final cluster, are removed from the pool, and the process continues until all remaining nodes fit within the size constraint.

The resulting set of clusters $\mathcal{C}$ forms a complete partition of the asset universe,
\begin{equation}
\bigsqcup_{m=1}^M \mathcal{C}_m = \{1,\dots,n\}, \qquad
|\mathcal{C}_m| \le Q_{\max} \ \ \forall m,
\end{equation}

guaranteeing that each cluster defines a QUBO subproblem embeddable on the target quantum hardware.

\begin{algorithm}[t]
\caption{Hardware-Aware Clustering (\textsc{GetClusters})}
\label{alg:get_clusters}
\DontPrintSemicolon

\KwIn{Correlation matrix \(\mathscr{C}\); hardware qubit limit \(Q_{\max}\)}
\KwOut{Clusters \(\{\mathcal{C}_1,\dots,\mathcal{C}_M\}\) with \(|\mathcal{C}_m| \le Q_{\max}\)}

\textbf{RMT denoising:} Decompose \(\mathscr{C}\) into
\(\mathscr{C}_{\mathrm{noise}}, \mathscr{C}_{\star}, \mathscr{C}_{\mathrm{global}}\) using
Marchenko-Pastur spectral separation\;

\textbf{Initial clustering~\cite{dcmppln2024}:} Apply a community-detection method (e.g.\ Louvain) to \(\mathscr{C}_{\star}\) to obtain initial communities \(\{\mathcal{C}^{(0)}_u\}\)\;

\ForEach{community \(\mathcal{C}^{(0)}_u\)}{

    \eIf{\(|\mathcal{C}^{(0)}_u| \le Q_{\max}\)}{
        Keep \(\mathcal{C}^{(0)}_u\) unchanged and append it to the output set\;
    }{
        \textbf{Initialize:} \(R \leftarrow \mathcal{C}^{(0)}_u\)\;

        \While{\(R \neq \emptyset\)}{

            \If{\(|R| \le Q_{\max}\)}{
                Append \(R\) as a final cluster; break\;
            }
            
            - Compute the restricted matrix \(\mathscr{C}_R = \mathscr{C}[R,R]\)\;

            - Compute degrees
            \(d_i = \sum_j |(\mathscr{C}_R)_{ij}| - 1\, \forall i \in R\)\;

            - Select seed
            \(i^\star = \arg\max_i d_i\)\;

            - Compute similarities
            \(s_j = |(\mathscr{C}_R)_{i^\star j}|\)\;

            - Let \(J\) be the indices of the \(Q_{\max}\) largest \(s_j\)\;

            - Form a new cluster
            \(\mathcal{C}_{\mathrm{new}} = \{ R[j] : j \in J \}\)\;

            - Append \(\mathcal{C}_{\mathrm{new}}\) to the output set\;

            - Update \(R \leftarrow R \setminus \mathcal{C}_{\mathrm{new}}\)\;
        }
    }
}
 
\Return{Final set of clusters \(\{\mathcal{C}_1,\dots,\mathcal{C}_M\}\)}\;

\end{algorithm}

\subsection{Trapped-ion hardware}
\label{sec:ion_trap_hardware}

The experiments reported here were carried out on modern trapped-ion quantum processors. This hardware platform is particularly well suited for implementing dense Ising or QUBO Hamiltonians due to its long intrinsic qubit coherence times and its ability to realize highly reconfigurable, near–all-to-all two-qubit interactions mediated by collective motional modes. These characteristics allow dense coupling graphs to be implemented directly, substantially reducing the need for SWAP operations that typically inflate circuit depth on devices with limited connectivity.

Our primary results were obtained using IonQ’s Forte and Forte Enterprise quantum processing units (QPUs)~\cite{Chen2024-co}, each of which utilize 36 qubits encoded in the hyperfine ground states of trapped $^{171}\text{Yb}^+$ ions, while selected experiments were performed on a larger, IonQ-designed Barium development system similar to the forthcoming IonQ Tempo line.
This system utilizes up to 64 Barium qubits in a long-chain, steered-beam architecture to extend the capabilities of Forte-class processors. Across all devices, ions are produced via laser ablation and selective ionization and are confined within compact, integrated vacuum packages utilizing surface linear Paul traps. Universal control is implemented through two-photon Raman transitions driven by $\SI{355}{nm}$ laser pulses for the Ytterbium systems and $\SI{532}{nm}$ laser pulses for the Barium system, enabling a gate set composed of arbitrary single-qubit rotations and entangling $ZZ$ gates. 

These QPUs leverage advanced optical control systems based on acousto-optic deflectors (AODs) to enable independent beam steering to individual ions, significantly reducing beam alignment errors~\cite{Kim:2008ApOpt,Pogorelov:2021PRXQ}, and are supported by automated calibration software to ensure scalable operation. Regarding performance, the IonQ Forte and Forte Enterprise systems demonstrated median direct randomized benchmarking fidelities for two-qubit gates of $99.3\%$ and $99.5\%$, respectively, with typical gate durations of $950~\mu\text{s}$. Single-qubit gate fidelities were approximately $99.98\%$ with operation times of $130~\mu\text{s}$. Data processing protocols across the systems include sample debiasing, achieved by grouping the total number of shots into batches of 25 and averaging over these groups~\cite{maksymov2023}. Furthermore, the 64-Qubit Barium development system incorporates leakage checks to discard samples where quantum states have interacted with the environment. Collectively, these architectures support consistent high gate fidelities, making them well-suited for quantum optimization experiments featuring dense interaction structures.

\subsection{Quantum optimizer}
\label{sec:quantum-optimizer}

Bias-field digitized counterdiabatic quantum optimization (BF-DCQO) is a recent algorithm proposed to tackle optimization problems encoded as Ising problems \cite{cadavid2025, Romero2025, chandarana2025}. BF-DCQO builds up on an adiabatic path from a biased initial Hamiltonian $H_i=\sum_j \left( -X_j + h^b_j Z_j \right)$ to the target Hamiltonian $H_f$, i.e. $H_{ad} = (1-\lambda) H_i + \lambda H_f$, where $\lambda$ is a scheduling function satisfying $\lambda(0)=0$, $\lambda(T)=1$, and $T$ being the evolution time. Specifically, in all experiments we use the smooth schedule $\lambda(t) = \sin^2\!\left[\frac{\pi}{2}\sin^2\!\left(\frac{\pi t}{2T}\right)\right]$, whose first derivative vanishes at \(t=0\) and \(t=T\).  Doing arbitrary long evolution times is not feasible experimentally due to decoherence. Then, a counterdiabatic (CD) protocol is needed to overcome this issue, which modifies the adiabatic path as $H = H_{ad} + \dot{\lambda} A_{\lambda}$. We choose $A_{\lambda}$ as the first order nested commutator expansion~\cite{claeys_nc}, namely $A_\lambda = i \alpha(\lambda) \comm{H_i}{H_f}$, where $\alpha(\lambda) = - \norm{\comm{H_i}{H_f}}^2 / \norm{\comm{H_{ad}}{\comm{H_i}{H_f}}}^2$. Throughout, \(\norm{\cdot}\) denotes the trace norm.

For short evolution times, the magnitude of the CD term dominates over the adiabatic one, i.e. $\norm{\dot{\lambda} A_\lambda} \gg \norm{H_{ad}}$, and the time evolution operator $\mathcal{U}$ is effectively generated by $H\approx\dot{\lambda} A_\lambda$. Doing this effective time evolution in a digital quantum computer is known as digitized counterdiabatic quantum optimization (DCQO) in the impulse regime~\cite{PhysRevApplied.22.054037}. The digitization of the time evolution operator can be done through Trotterization, yielding 
\begin{equation}
    \mathcal{U}_\text{dig} = \prod_{k=1}^{n_\text{steps}} \prod_{j=1}^{n_\text{terms}} e^{-i \Delta t \, H_j(k\Delta t)},
\end{equation}
where $T=n_\text{steps} \Delta t$ and the Hamiltonian has been expanded as $H=\sum_{j=1}^{n_\text{terms}} H_j$, see Algorithm~\ref{alg:dcqo}. To further reduce circuit depth and accumulated noise in the digital implementation, we employ a pruning procedure at the compilation level. Pruning consists of discarding gates whose rotation angles fall below a fixed threshold $\theta_\text{cut-off}$, as these correspond to weak contributions to the implemented evolution. In practice, this acts as a controlled approximation that lowers hardware overhead while preserving the dominant structure of the cost Hamiltonian. 

The circuits are built using Qiskit~\cite{qiskit}. Once compiled, BF-DCQO applies DCQO iteratively by modifying the bias fields $h^b_j$ from $H_i$, favoring a certain orientation of the qubits, see Algorithm~\ref{alg:bf_dcqo}. Furthermore, at each iteration we select the $n_{l}$ lowest-energy samples and perform five sweeps of zero-temperature simulated annealing as post-processing to correct for bit-flip errors while getting better samples. We then update the bias fields using the empirical spin orientations of the measured distribution $h_i^{(b)} \leftarrow -\langle \sigma_i^z \rangle$, where \(\langle \sigma_i^z \rangle\) is computed from the subset of post-processed $n_l$ samples. In this work, we limit ourselves to $n_l = 10$. Additionally, further problem-specific post-processing is admitted on the generated samples, as will be described in the next section. 

\begin{algorithm}[t]
\caption{Digitized Counterdiabatic Evolution (DCQO)}
\label{alg:dcqo}
\DontPrintSemicolon

\KwIn{Initial Hamiltonian \(H_i\), target Hamiltonian \(H_f\), schedule \(\lambda(t)\), time step \(\Delta t\), threshold \(\theta_\text{cut-off}\)}
\KwOut{Digitized unitary evolution operator \(\mathcal{U}_{\mathrm{dig}}\)}

Compute the counterdiabatic term \(A_{\lambda} = i \alpha(\lambda) [H_i, H_f]\)\;
Initialize \(\mathcal{U}_{\mathrm{dig}} \leftarrow I\)\;

\For{\(k = 1 \dots n_{\mathrm{steps}}\)}{
    -Let $\lambda_k=\lambda(k \Delta t)$ \;
    -Evaluate \(H_k = \dot{\lambda}_k A_{\lambda_k}\)\;
    -Decompose \(H_k = \sum_j \, r_{j,k} P_{j,k}\), with $P_{j,k}$ a Pauli product\;
    -Compute the Trotterized step
    \[
       \mathcal{U}_k = \prod_{j\in\mathcal{J}} \exp\!\left(-i\,\Delta t \, r_{j,k} P_{j,k}\right)
    \]\;
    where $\mathcal{J}$ is the set of indices for which $\abs{\Delta t \, r_{j,k}} > \theta_\text{cut-off}$ \;
    -Update \(\mathcal{U}_{\mathrm{dig}} \leftarrow \mathcal{U}_k\, \mathcal{U}_{\mathrm{dig}}\)\;
}

\Return{\(\mathcal{U}_{\mathrm{dig}}\)}\;

\end{algorithm}

\begin{algorithm}[t]
\caption{Bias-Field Digitized Counterdiabatic Quantum Optimization (BF-DCQO)}
\label{alg:bf_dcqo}
\DontPrintSemicolon

\KwIn{Target Hamiltonian \(H_f\); initial bias fields \(h^{(b)}_j\); number of iterations \(R\), number of low energy samples to keep $n_l$}
\KwOut{Best sampled bitstring}

\For{\(r = 1 \dots R\)}{

    -Construct biased driver
    \[
        H_i^{(r)} = \sum_j \left(-X_j + h^{(b)}_{j,r} Z_j\right)
    \]\;

    -Execute DCQO using Algorithm~\ref{alg:dcqo} and obtain measurement samples\;

    -Compute classical energies of all measured bitstrings under \(H_f\)\;

    -Sort samples by increasing energy\;

    -Post-select the $n_l$ lowest-energy samples and estimate \(\langle \sigma_j^z \rangle\) \;
    -Set \(h^{(b)}_{j,r+1} \leftarrow -\langle \sigma_j^z \rangle\)\;
}

\Return{Distributions of bitstrings from the iterations}\;

\end{algorithm}

\subsection{Post-processing}
\label{subsec:local_search_cardinality}

The use of classical post-processing is motivated by two practical considerations. First, the quantum circuits used in this work do not explicitly preserve the cardinality constraint, and sampling noise further spreads the measurement outcomes around the target Hamming weight. As a result, low-energy samples produced by BF-DCQO typically concentrate near, but not exactly on, configurations with the desired cardinality. This first phase of the post-processing therefore performs a deterministic repair that projects these near-feasible solutions onto the constraint manifold in a controlled manner.

Second, once feasibility is restored, the repaired configuration can be viewed as a high-quality starting point for a short local exploration. We perform a lightweight neighborhood search restricted to fixed Hamming weight. This reflects the broader working hypothesis that near-term quantum optimizers can be effective at identifying promising regions of the search space, while classical heuristics remain well-suited to perform fine-grained improvements~\cite{chandarana2025hybridsequentialquantumcomputing}. We present this two-phase approach in Algorithm~\ref{alg:local_search_cardinality}.

\textit{Gradient-based Hamming weight recovery.––} Let $w(x) = \sum_{i=1}^n x_i$ denote the Hamming weight of $x$, and let $\nabla f(x)$ be the continuous gradient of the quadratic form
\begin{equation}
  \nabla f(x) = 2Qx + q.
\end{equation}
If $w(x) = K$, we skip this phase. Otherwise:

\begin{itemize}
  \item If $w(x) > K$, we must deactivate $(w(x)-K)$ assets.
        Among all indices $i$ with $x_i = 1$, we sort them by
        decreasing $\nabla f(x)_i$ and flip to $0$ the first
        $w(x)-K$ indices. Intuitively, large positive gradient entries
        indicate variables that are most harmful to keep active.
  \item If $w(x) < K$, we must activate $(K-w(x))$ assets.
        Among all indices $i$ with $x_i = 0$, we sort them by
        increasing $\nabla f(x)_i$ and flip to $1$ the first
        $K-w(x)$ indices. Small (or negative) gradient entries
        indicate variables that are beneficial to activate.
\end{itemize}

\begin{algorithm}[t]
\caption{Two-phase local search with cardinality constraint}
\label{alg:local_search_cardinality}
\DontPrintSemicolon
\KwIn{Initial $x \in \{0,1\}^n$; QUBO $(Q,q,c)$; target weight $K$; max iters $T_{\max}$; patience $P$.}
\KwOut{Locally improved $\tilde{x}$ with $\sum_i \tilde{x}_i = K$.}

$f_{\text{init}} \gets x^\top Q x + q^\top x + c$; \quad
$w \gets \sum_i x_i$ \;

\BlankLine
\textbf{Phase 1: cardinality repair}\;
\If{$w \neq K$}{
  $g \gets 2Qx + q$ \;
  \lIf{$w > K$}{
    $S \gets \{i : x_i = 1\}$ sorted by $g_i$ in descending order;
    flip $x_i \gets 0$ for first $(w-K)$ indices in $S$
  }
  \lElse{
    $S \gets \{i : x_i = 0\}$ sorted by $g_i$ in ascending order;
    flip $x_i \gets 1$ for first $(K-w)$ indices in $S$
  }
}
$f_{\text{p1}} \gets x^\top Q x + q^\top x + c$; \quad
$\Delta_{\text{p1}} \gets f_{\text{init}} - f_{\text{p1}}$ \;

\BlankLine
\textbf{Phase 2: swap local search}\;
$f \gets f_{\text{p1}}$; \quad
$t \gets 0$; \quad
$p \gets 0$ \;

$S_0 \gets \{i : x_i = 0\}$; \quad
$S_1 \gets \{j : x_j = 1\}$ \;
$\text{maxSwaps} \gets \min\bigl(100, |S_0||S_1|\bigr)$ \;

\While{$t < T_{\max}$ \textbf{ and } $p < P$}{
  $\text{improved} \gets \text{false}$; \quad
  $s \gets 0$ \;

  randomly permute $S_0$ and $S_1$ \;

  \For{$i \in S_0$}{
    \If{$\text{improved}$ or $s \ge \text{maxSwaps}$}{\textbf{break}}
    \For{$j \in S_1$}{
      \If{$s \ge \text{maxSwaps}$}{\textbf{break}}
      $x_i \gets 1$, $x_j \gets 0$; \quad
      $f' \gets x^\top Q x + q^\top x + c$; \quad
      $s \gets s+1$ \;

      \If{$f' < f$}{
        $f \gets f'$; \quad
        $S_0 \gets \{k : x_k = 0\}$; \quad
        $S_1 \gets \{k : x_k = 1\}$ \;
        $\text{improved} \gets \text{true}$; \quad
        $p \gets 0$; \quad
        \textbf{break}
      }
      \Else{
        $x_i \gets 0$, $x_j \gets 1$
      }
    }
  }

  \If{\textbf{not} $\text{improved}$}{ $p \gets p+1$ }
  $t \gets t+1$ \;
}

\Return{$\tilde{x} \gets x$} \;
\end{algorithm}

This phase yields a bit string $x$ that satisfies $w(x) = K$ with
a single pass of gradient-based flips.

\textit{First-improvement swap-based local search.––} Starting from the repaired configuration $x$ of weight $K$, we perform a local search over the cardinality-preserving neighborhood defined by all pairwise swaps $(i,j)$ with $x_i = 0$ and $x_j = 1$:
\[
  x' = x + e_i - e_j,
\]
where $e_i$ is the $i$-th unit vector. In each iteration, we randomly
permute the sets of indices with $x_i=0$ and $x_j=1$ and scan swap pairs in this randomized order. The first swap that yields a strict improvement $f(x') < f(x)$ is accepted immediately (first-improvement rule), and the index sets are updated accordingly. To control runtime, we limit the number of swap evaluations per iteration to $100$ and terminate the search if either a maximum number of iterations is reached or no improving swap is found for a prescribed number of consecutive iterations. The procedure returns the locally optimal (w.r.t.\ swap moves) bit string of weight $K$, together with bookkeeping information such as the number of iterations and the improvements achieved in each phase.

\section{Results}
\label{sec:experimental_results}

We benchmark the hardware-aware quantum optimization pipeline described in \seclabel{sec:portfolio_optimization} and \seclabel{sec:methods} on real data. Our goal is to assess how the decomposition adapts to different hardware qubit limits, and how the BF-DCQO plus local-search combination performs on the resulting portfolio optimization instances. In particular, we focus on the end-to-end executability of the pipeline, starting from a dense $250$-asset objective, proceeding through hardware-sized Ising subproblems, and culminating in reconstructed global portfolios that satisfy the fixed-cardinality constraint.

\begin{table}[t!]
\centering
\caption{Cluster statistics for the hardware-aware decomposition of a
250-asset universe under different qubit limits \(Q_{\max}\). In bold font the system sizes that were tackled on quantum hardware.}
\label{tab:cluster_stats}
\begin{tabular}{lcc}
\hline\hline
Hardware & \#\,Clusters & Cluster sizes \\
\hline
36 qubits & 14 & [2,8,\textbf{36},5,10,13,10,1,26,24,\textbf{36},19,\textbf{36},24] \\
64 qubits & 11 & [2,8,\textbf{41},10,13,10,1,26,24,\textbf{55},\textbf{60}] \\
\hline\hline
\end{tabular}
\end{table}

\begin{figure}[t!]
    \centering
    \includegraphics[width=1\linewidth]{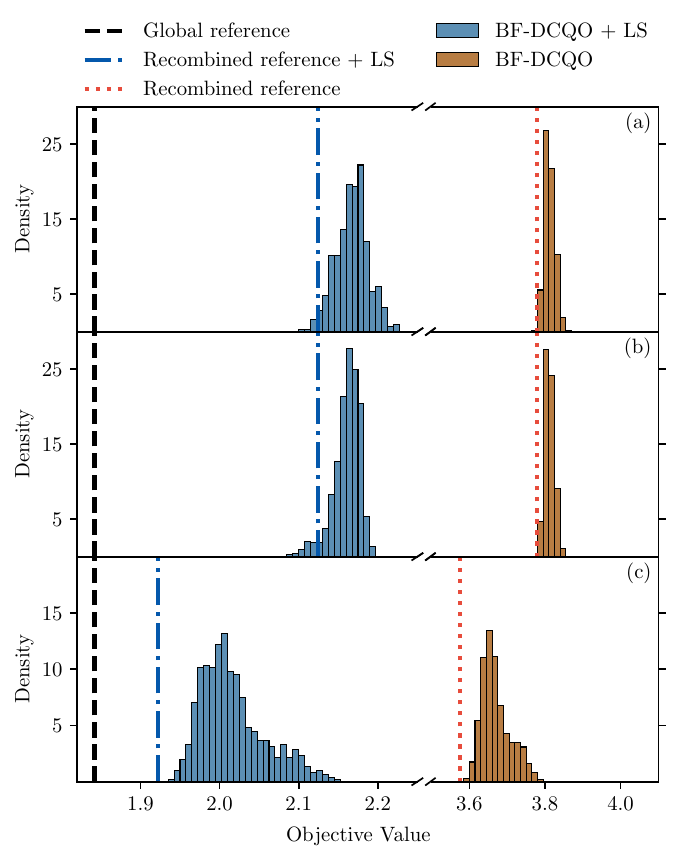}
    \caption{Energy distribution of global candidate portfolios under different hardware and pruning configurations. From top to bottom, the panels correspond to (a) the 36-qubit decomposition with high pruning on IonQ Forte, (b) the 36-qubit decomposition with medium pruning on IonQ Forte, and (c) the 64-qubit decomposition with medium pruning on the IonQ 64-qubit Barium development system. Orange histograms show the energy distribution obtained by running BF-DCQO on the largest subproblem, applying the two-phase local search at the cluster level and then, recombining into global candidates. Blue histograms show the corresponding distributions after applying the two-phase local-search post-processing. Red dotted vertical lines indicate the energy of the recombined optimal solution obtained by merging the individually optimal subproblem solutions, while blue dash-dotted lines denote this recombined optimum after local search. The black dashed lines mark the global optimum of the full problem, obtained by the Gurobi solver.}
    \label{fig:portfolios-energy-distribution}
\end{figure}

\begin{figure}[htb]
    \centering
    \includegraphics[width=1\linewidth]{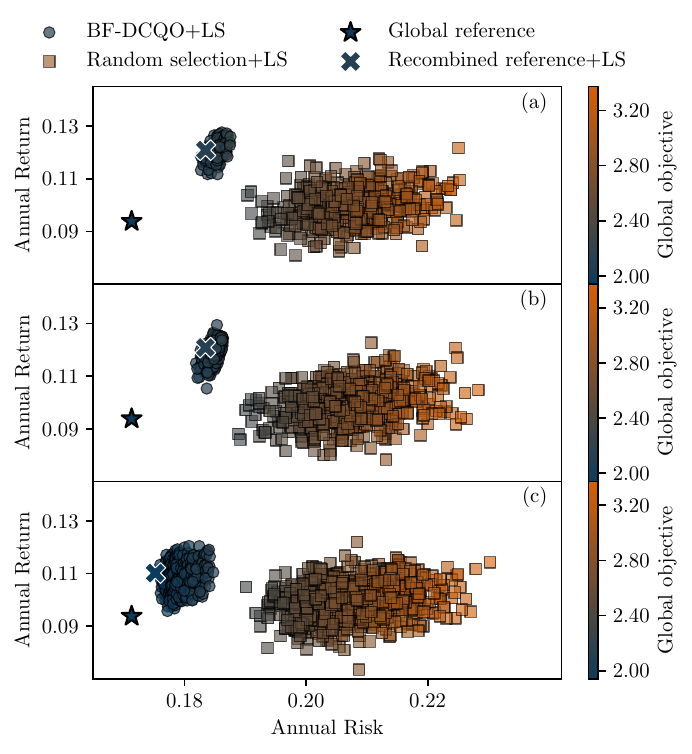}
    \caption{Return-risk distribution of global candidate portfolios generated by the clustered BF-DCQO pipeline under different hardware and pruning configurations. From top to bottom, the panels correspond to (a) the 36-qubit decomposition with high pruning, (b) the 36-qubit decomposition with medium pruning, and (c) the 64-qubit decomposition with medium pruning. Square markers denote portfolios obtained from random initialization followed by classical post-processing, while circular markers represent post-processed candidates. The global reference portfolio is indicated by a red star, and the recombined reference portfolio is shown as a blue ×. The color scale encodes the objective value of each portfolio. The post-processing budget is intentionally small (Sec.~\ref{subsec:local_search_cardinality}) so that differences primarily reflect the quality of the initial candidate pools.}
    \label{fig:portfolios-return-risk}
\end{figure}

We construct the universe of \(n=250\) assets from the S\&P~500 index. For each constituent, we use daily closing prices over the period \texttt{2020-01-02} to \texttt{2025-08-22}, corresponding to \(1418\) trading days. Daily log-returns are then computed as \(r_t = \log(p_t/p_{t-1})\). From the resulting return matrix, we select the \(250\) assets with the most complete history, leading to a final data set with dimensions approximately \((1418,250)\). From these returns, we compute the sample mean vector, covariance matrix, and the empirical correlation matrix used in the decomposition pipeline. We take $K=n/2=125$ for the global cardinality constraint.

We consider two target devices with different limits: \(Q_{\max}=36\) and \(Q_{\max}=64\) qubits, corresponding to IonQ Forte and the 64-qubit Barium development system. In both cases, the full $250$-asset universe is partitioned into independent subproblems whose sizes do not exceed the corresponding \(Q_{\max}\). Following Algorithm~\ref{alg:get_clusters}, for the 36-qubit setting we obtain $14$ clusters with sizes  \([2,8,36,5,10,13,10,1,26,24,36,19,36,24]\), yielding three subproblems of the maximal size $36$. For the $64$-qubit setting the same universe is grouped into $11$ clusters of sizes \([2,8,41,10,13,10,1,26,24,55,60]\), with the largest subproblems containing $41$, $55$, and $60$ assets. This is summarized in \tablabel{tab:cluster_stats}. As expected, increasing the qubit budget reduces the number of clusters and allows for larger instances to be treated directly on quantum hardware. 

For each cluster we construct the cost function \(f_\text{subproblem}(x)\) using $K_\text{subproblem}=n_\text{subproblem}/2$ for the cardinality constraint. We map it to an Ising Hamiltonian \(H_f\), and run the BF-DCQO algorithm. All cluster problems are solved with the same quantum hyperparameters: two Trotter steps with time step \(\Delta t = 0.1\), and $10$ iterations per tackled subproblem. For the IonQ Forte runs, we only tackle the three $36$ qubit subproblems using BF-DCQO with $4000$ shots per iteration. The remaining subproblems of smaller size are solved exactly using Gurobi~\cite{gurobi} via Qiskit~\cite{qiskit}. Similarly, we only tackle the $41,55,60$-qubit instances with the 64-qubit Barium development system using $4000$, $5000$ and $6000$ shots per iteration, respectively. We also define two levels of pruning for each case. For the $36$-qubit decomposition, medium pruning removes roughly $40\%$ of the gates following Algorithm~\ref{alg:dcqo}, whereas high pruning the $60\%$. On the other hand, for the $64$-qubit decomposition, we only implement medium pruning, which removes about $50\%-70\%$ of the gates. 

To construct approximate solutions for the global $250$-asset portfolio, we recombine cluster solutions by direct concatenation. For a fixed cluster ordering, a global bitstring is formed by placing the bits of each cluster-level candidate into their corresponding asset positions, yielding a full-length portfolio configuration. Based on this recombination rule, we build a candidate pool of size $1000$ by sampling and post-processing low-energy cluster solutions. Specifically, for each cluster tackled with BF-DCQO, we collect the best bitstring from each iteration, adding up to a total of $10$ bitstrings per cluster. These bitstrings are then post-processed at the cluster-level following the two-phase local search procedure of \seclabel{subsec:local_search_cardinality}, using a maximum of $100$ iterations and an early-stop after one iteration without improvement. The raw distributions for each subproblem are shown in Appendix~\ref{app:raw-dist}. In this way, we generate $1000$ global independent bitstrings, providing a diverse set of candidate portfolios concentrated in low-energy regions of the global objective. Then, these global candidates are further refined using the two-phase local search.

Figure~\ref{fig:portfolios-energy-distribution} shows the distribution of objective values of global candidate portfolios generated by recombining low-energy solutions of the clustered subproblems. A comparison of the three panels shows that increasing the available qubit budget from \(Q_{\max}=36\) to \(Q_{\max}=64\) systematically improves the best post-processed objective values obtained from the recombined candidate pool. The \(64\)-qubit decomposition yields post-processed solutions that are closer to the global reference optimum than those obtained in the \(36\)-qubit cases. At fixed hardware size (\(36\) qubits), medium pruning produces a slightly but not significantly lower-energy distribution than high pruning. This suggests that lower-resource Hamiltonian representations may remain competitive in practice, particularly on near-term devices where reduced circuit complexity can partially offset the loss of model fidelity.

Importantly, Fig.~\ref{fig:portfolios-energy-distribution} indicates that, in several cases, the best post-processed portfolios achieve objective values lower than those of the post-processed recombined reference, which is obtained by merging the individually optimal subproblem solutions and applying local search on it. This behavior highlights the nontrivial role of global local search: candidate portfolios that are suboptimal prior to post-processing can be steered toward lower-energy configurations once the global cardinality constraint is enforced and inter-cluster swap moves are allowed. As a result, the recombined reference should be viewed as a baseline rather than an upper bound on achievable solution quality within the hybrid pipeline.

Despite the substantial improvement introduced by local search, the post-processed distributions do not collapse onto the global optimum, even in the largest (\(64\)-qubit) hardware setting. This residual gap indicates that the dominant limitations arise from the approximate nature of the decomposition. Nevertheless, the observed monotonic improvement with increasing qubit budget demonstrates that the hardware-aware decomposition adapts smoothly to larger devices and that larger embeddable subproblems translate directly into higher-quality global portfolios within the same hybrid pipeline.

Figure~\ref{fig:portfolios-return-risk} represents the global candidate portfolios in the annualized risk-return plane. While the QUBO/Ising objective ranks portfolios by a single scalar value, it does not reveal how solutions trade off financial risk against expected return. Showing candidate solutions in the risk-return plane, therefore, allows one to assess the financial structure and diversity of the solution set produced by the hybrid pipeline.

Across all configurations, BF-DCQO with local search (LS) generates a broad distribution of feasible portfolios spanning a continuous range of risk and return levels. The presence of extended bands of solutions with comparable objective values indicates a high degree of near-degeneracy in the combinatorial objective, whereby portfolios with similar energies can exhibit meaningfully different financial characteristics. In particular, low-objective portfolios are not confined to extreme-risk or extreme-return regions, but instead populate the interior of the feasible domain, demonstrating that objective improvements are associated with financially interpretable trade-offs.

\section{Conclusions}
We presented a hardware-aware decomposition pipeline for fixed-cardinality portfolio optimization that enables end-to-end execution on today's and near-term gate-based quantum processors. The central design choice is to treat the executable qubit budget as an explicit constraint during decomposition, producing hardware-sized Ising subproblems and a practical path from empirical market data to feasible portfolios under realistic device limits.

On a 250-asset equity universe, we evaluated the full workflow under trapped-ion-motivated qubit budgets and executed the largest subproblems on IonQ hardware using BF-DCQO. We observe that increasing the executable subproblem size improves the final global portfolio quality after recombination and post-processing, consistent with reduced decomposition-induced approximation error as fewer cross-cluster interactions are omitted. At a fixed qubit budget, varying the pruning level yields similar end-to-end performance, indicating that moderate approximations of the implemented Hamiltonian can reduce circuit cost without necessarily degrading the final solution quality on current noisy hardware. These observations clarify a practical trade-off between circuit complexity and optimization performance in decomposition-based quantum workflows.

Several directions merit further study. Algorithmically, incorporating cross-cluster couplings more explicitly during recombination and expanding the neighborhood structure used in global refinement could further close the gap to the global optimum. From the hardware perspective, larger qubit counts and improved gate fidelities would directly translate into larger executable subproblems and fewer clusters, improving solution quality within the same pipeline. Finally, because subproblems can be solved independently, the workflow is naturally parallel and could leverage multiple homogeneous or heterogeneous QPUs. Overall, our results establish hardware-aware decomposition as a practical and scalable pathway for bringing quantum optimization methods to structured financial decision problems on today's and near-term quantum processors.

\begin{acknowledgments}
We thank Michael Wurster and Sebastian Wagner for their support on the Kipu Quantum Hub platform while running the experiments. We thank Coleman Collins, Neal Pisenti, and Ken Wright for discussions and their support. \textit{}
\end{acknowledgments}

\appendix

\section{QUBO/Ising formulation}
\label{appendix:ising-qubo-formulation}

The optimization of Eq.~(\ref{eq:min-problem}) contains a cardinality constraint. In this appendix, we show how to convert this constrained problem into an unconstrained one, and additionally, how to map it into an Ising Hamiltonian.

We define the QUBO cost function $f(x)$ as a penalized objective for the configurations that violate the constraint, namely
\begin{equation}
    f(x) = \frac{\gamma-1}{2}\mu^T x + \frac{\gamma+1}{2} x^T Cx + \Lambda \left(\sum_{i=1}^{n} x_i - K \right)^2,
\end{equation}
where $\Lambda$ is the Lagrange multiplier, quantifying the strength of the penalty for invalid configurations. Expanding the penalty term results in
\begin{align}
    f(x)&=\left(\frac{\gamma-1}{2} \mu^T - 2K\Lambda 1^T  \right)x + x^T \left( \frac{1+\gamma}{2}C-\Lambda 1\cdot1^T \right) x + K^2\Lambda \notag \\
    &=x^\top Q x + q^\top x + K^2\Lambda,
\end{align}

\begin{figure}[htb]
    \centering
    \includegraphics[width=1\linewidth]{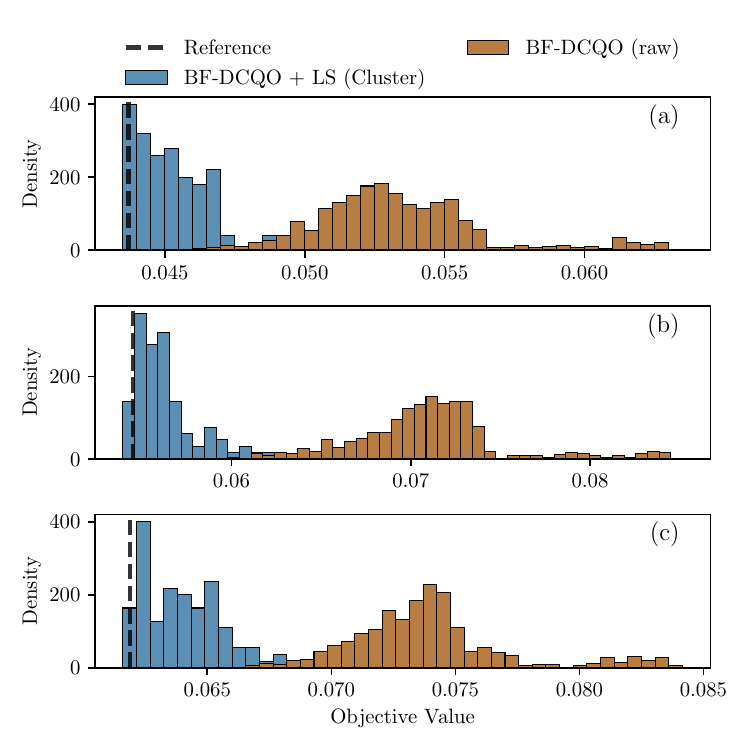}
    \caption{Superposed top-$10$ samples of each iteration of BF-DCQO on IonQ Forte using high pruning. We show the distributions before and after cluster-level local search for (a) the first $36$-qubit cluster, (b) the second $36$-qubit cluster and (c) the last $36$-qubit cluster.}
    \label{fig:app-36-h}
\end{figure}

\begin{figure}[htb]
    \centering
    \includegraphics[width=1\linewidth]{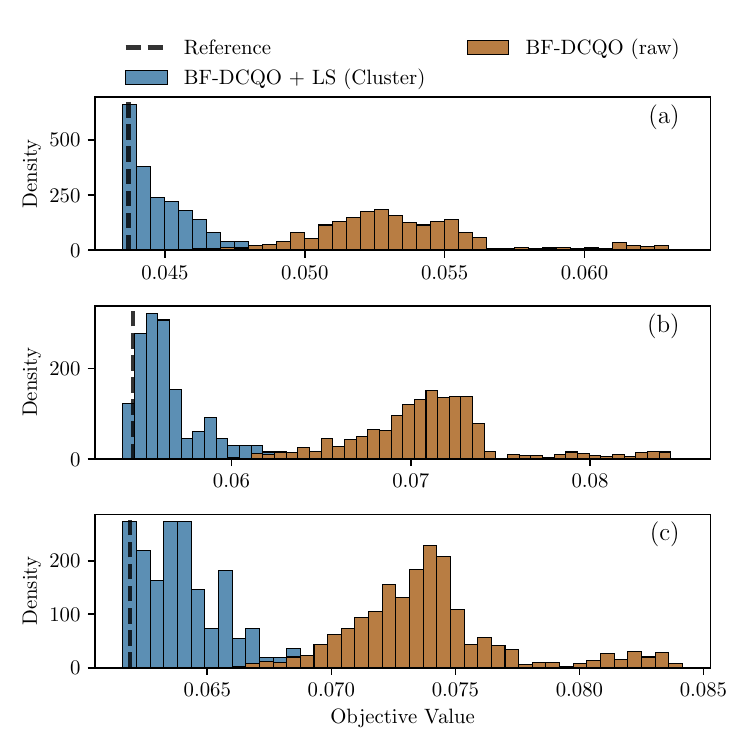}
    \caption{Superposed top-$10$ samples of each iteration of BF-DCQO on IonQ Forte using medium pruning. We show the distributions before and after cluster-level local search for (a) the first $36$-qubit cluster, (b) the second $36$-qubit cluster and (c) the last $36$-qubit cluster.}
    \label{fig:app-36-m}
\end{figure}

\begin{figure}[htb]
    \centering
    \includegraphics[width=1\linewidth]{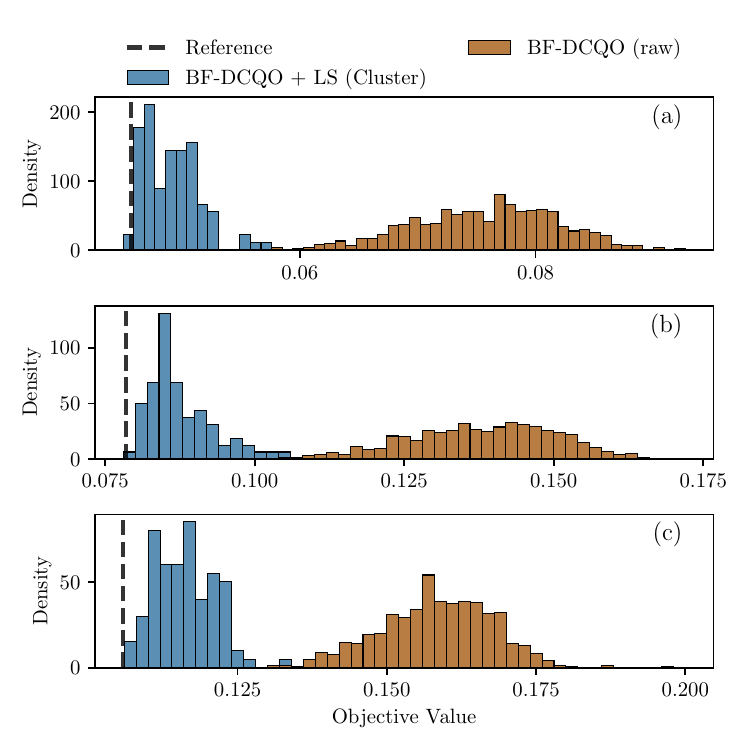}
    \caption{Superposed top-$10$ samples of each iteration of BF-DCQO on the 64-qubit Barium development system using medium pruning. We show the distributions before and after cluster-level local search for (a) the $41$-qubit cluster, (b) the $55$-qubit cluster and (c) the $60$-qubit cluster.}
    \label{fig:app-64-m}
\end{figure}

which is equivalent to Eq.~(\ref{eq:cost_fn}). Additionally, in this work, we pick the penalty coefficient as

\begin{equation}
\Lambda = 2 \cdot \max_{i \in \{1, \ldots, n\}} I_i,
\end{equation}

where $I_i = |q_i + Q_{ii}| + \sum_{j=1, \, j \neq i}^{n} |Q_{ij}|$. This ensures that constraint violations are always more costly than any potential benefit from the original objective. The factor of 2 provides a conservative safety margin, while the influence-based calculation makes the penalty adaptive to the problem's coefficient magnitudes.

In order write the cost function as a Hamiltonian $H_f$, whose ground state is the solution of the optimization problem, we perform the mapping $x\to\frac12(1-z)$, with $z=(z_1,\cdots,z_n)$ and $z_i\in[-1,1]$ the Pauli-Z operator for the $i$-th spin. After substitution, we obtain
\begin{align}
H_f&=\frac{1}{4}z^T Q z - \frac12 \left( 1^T Q + q^T \right)z + \frac14 1^TQ1 + \frac12 q^T 1 + K\Lambda \notag \\
&=h^T z + z^T J z + \text{const.},
\end{align}
which is equivalent to Eq.~(\ref{eq:cost_ising})

\section{Raw quantum distributions for the subproblems}
\label{app:raw-dist}

In this section, we show the distributions for the clusters of the $36$-qubit decompositions as well as the $64$-qubit ones, see Figs.~\ref{fig:app-36-h},\ref{fig:app-36-m} and \ref{fig:app-64-m}. The qualitative differences between the $36$- and $64$- qubit runs distributions can be explained by two main factors. First, on the hardware side, the device exhibited stronger noise, making it more difficult to skew the distributions effectively. In addition, the entangling gate budget was kept the same for both the 36-qubit and 64-qubit runs. As a result, the larger system suffered from a reduced solution quality. Second, algorithmically, the bias update for the 36-qubit cases was computed using only $10$ states, whereas for the 64-qubit case, we used $100$ states. This choice was made to prevent overly strong early biases, given the larger Hilbert space. While this helps avoid trapping effects as system size increases, it can also cause the distribution to have a larger variance, especially in the presence of noise.

\bibliography{reference.bib}

\end{document}